\documentclass[journal]{IEEEtran}
\ifCLASSINFOpdf
\else
\fi
\usepackage[dvipsnames]{xcolor}
\usepackage[normalem]{ulem}
\usepackage{color,soul}
\usepackage{enumitem}
\usepackage{comment}
\usepackage{longtable}
\usepackage{graphicx}
\usepackage[font=normalsize]{caption}
\usepackage{chemformula}
\usepackage{caption}
\usepackage{booktabs}
\DeclareCaptionLabelSeparator{tableNewline}{\par}
\usepackage{caption}
\usepackage{subcaption}
\usepackage[font={small},skip=10pt]{caption}
\usepackage{booktabs}
\usepackage[english]{babel}
\usepackage{multicol}
\usepackage{amsmath}
\usepackage{fancyhdr}
\usepackage[section]{placeins}
\usepackage{multirow}
\usepackage{flushend}
\usepackage{tabularx}
\usepackage{adjustbox}
\usepackage{float}
\usepackage{makecell}
\usepackage{booktabs}
\usepackage{array}
\usepackage{hyperref}
\usepackage{tikz}
\usepackage{quantikz}
\hypersetup{
  colorlinks=true,
  allcolors=blue,
  allbordercolors={blue!50!black}}
\usepackage[noadjust]{cite}
\usepackage[english]{babel}
\usepackage{graphicx}
\newcommand{\bluecite}[1]{\textcolor{blue}{\cite{#1}}}
\newcommand{\textblue}[1]{\textcolor{blue}{#1}}

\begin{document}
\bstctlcite{IEEEexample:BSTcontrol}
\setlength{\parskip}{0pt}
\title{Molecular Quantum (MolQ) Communication Channel in the Gut-Brain Axis Synapse}
\author{Bitop Maitra,~\IEEEmembership{Student Member,~IEEE}
        and~Ozgur~B.~Akan,~\IEEEmembership{Fellow,~IEEE}
\thanks{The authors are with the Center for neXt-generation Communications (CXC), Department of Electrical and Electronics Engineering, Ko\c{c} University, Istanbul 34450, Turkey (e-mail: \{bmaitra23, akan\}@ku.edu.tr).}
\thanks{O. B. Akan is also with the Internet of Everything (IoE) Group, Electrical Engineering Division, Department of Engineering, University of Cambridge, Cambridge CB3 0FA, UK (email: oba21@cam.ac.uk).}
       
\thanks{This work was supported by the AXA Research Fund (AXA Chair for Internet of Everything at Ko\c{c} University).}
\vspace{-1.2mm}
}

\maketitle

\begin{abstract}
The gut-brain axis is the communication link between the gut and the brain. Although it is known that the gut-brain axis plays a pivotal role in homeostasis, its overall mechanism is still not known. 
However, for neural synapses, classical molecular communication is described by the formation of ligand-receptor complexes, which leads to the opening of ion channels.
Moreover, there are some conditions that need to be fulfilled before the opening of the ion channel. 
In this study, we consider the gut-brain axis, where neurotransmitters diffuse through the synaptic cleft, considering molecular communication. 
On the vagus nerve (VN) membrane, i.e., the post-synaptic membrane of the synapse, it undergoes a quantum communication (QC), which initiates the opening of the ion channel, thus initiating the communication signal from the gut to the brain. 
It evolves a new paradigm of communication approach, Molecular Quantum (MolQ) communication. 
Based on the QC model, we theoretically analyze the output states, and QC is simulated considering the incoming neurotransmitter’s concentration and validated by analyzing the entropy and the mutual information of the input, i.e., neurotransmitter's concentration, and output, i.e., ion channel opening.

\end{abstract}

\begin{IEEEkeywords}
Gut-brain axis, Molecular communication, Quantum communication, Synaptic ion channel. 

\end{IEEEkeywords}

\renewcommand{\figurename}{Fig.}

\section{Introduction}
\label{sec:Intro}
\begin{figure}[t!]
    \centering
    \includegraphics[width=0.3\textwidth]{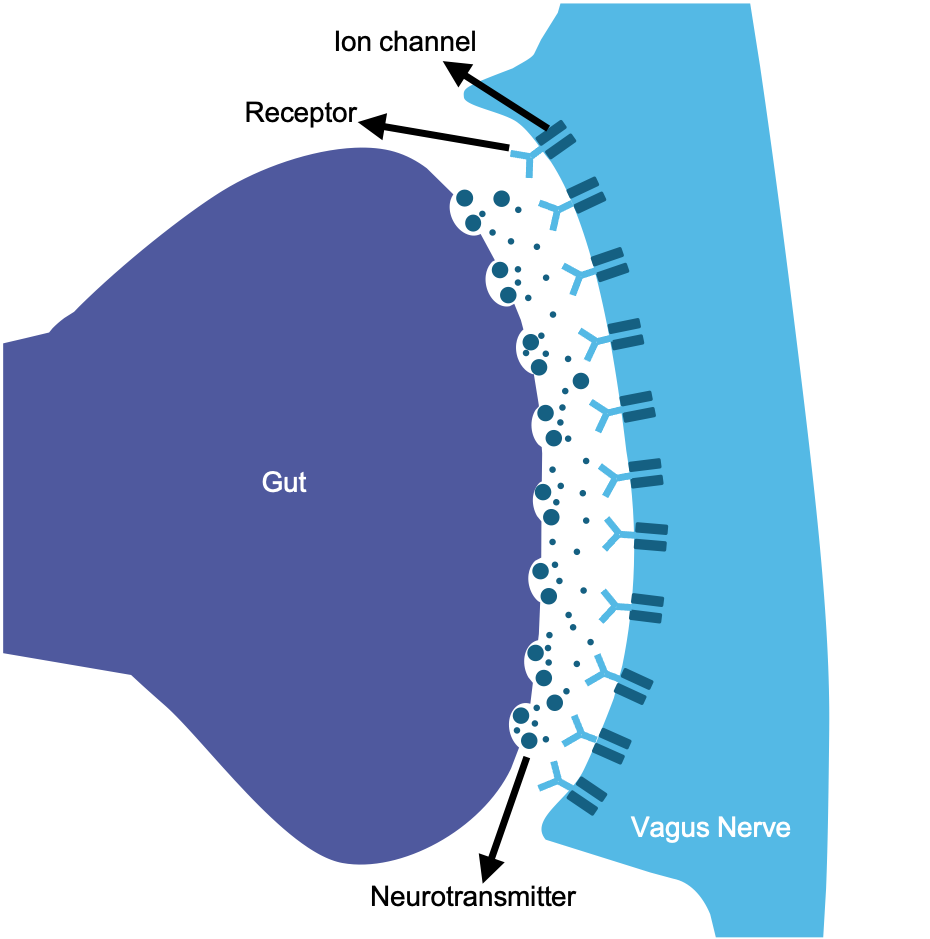}
    \caption{Gut-brain axis synapse.}
    \label{fig:GBAx}
\end{figure}

\IEEEPARstart {G}{ut-Brain Axis} (GBAx) is a bidirectional bioelectrochemical communication link between the gastrointestinal tract (GI-tract) or the Enteric Nervous System (ENS), often called the \textit{Second Brain} and the Central Nervous System (CNS), as shown in Fig. \ref{fig:GBAx}. 
GBAx is a complex communication link that includes neural, endocrine, immune, and humoral pathways.
The Vagus Nerve (VN), a part of the peripheral nervous system (PNS) is crucial in transmitting information from the gut to the brain and acts as a modulator \bluecite{breit2018vagus}. 
The key components of GBAx are gut microbiota (GM), the CNS, and the Neuro-Endocrine System (NES) \bluecite{carabotti2015gut}, where GM is the carrier of gut information, i.e., any alteration in GM can lead to significant perturbation in usual information transmission and affect the mind and whole body.
GM can influence the cognitive system: GBAx is the channel for maintaining homeostasis, affecting cognitive functions like memory, perception, hunger, etc.
Furthermore, a healthy gut bacteria can directly influence cognitive health and do disease prevention \bluecite{fekete2024exploring}, whereas any alteration in the GM can alter its composition, known as gut dysbiosis and lead to GBAx dysfunction and neurodegeneration like Parkinson's and Alzheimer's disease \bluecite{zheng2023understanding}, and mental health issues like anxiety and depression \bluecite{clapp2017gut}.
Therefore, GM is an inherently important factor in controlling the nervous system. 
The information of GM is sent to the brain by producing different neurotransmitters like acetylcholine \bluecite{chen2022neurotransmitter}, signalling molecules like Serotonin \bluecite{mawe2013serotonin} via interacting with post synaptic transmitter specific receptors.
Even though it has been speculated that GBAx is the most crucial communication pathway to control neuropsychiatry, less research has been conducted on it.
In this study, we will focus on the communication between the gut and the vagus nerve, where mostly the neurotransmitters are released vesicle-wise from the gut membrane, i.e., the pre-synaptic membrane (PSM) in the synaptic cleft of the GBAx synapse \bluecite{berghe2007spatial}. 
Those released molecules or particles travel through the synaptic cleft to reach the vagus nerve membrane (VNM), which can be called a post-synaptic membrane (POSM), by the process of molecular communication.

\begin{figure*}[t!]
    \centering
    \includegraphics[width=0.95\textwidth]{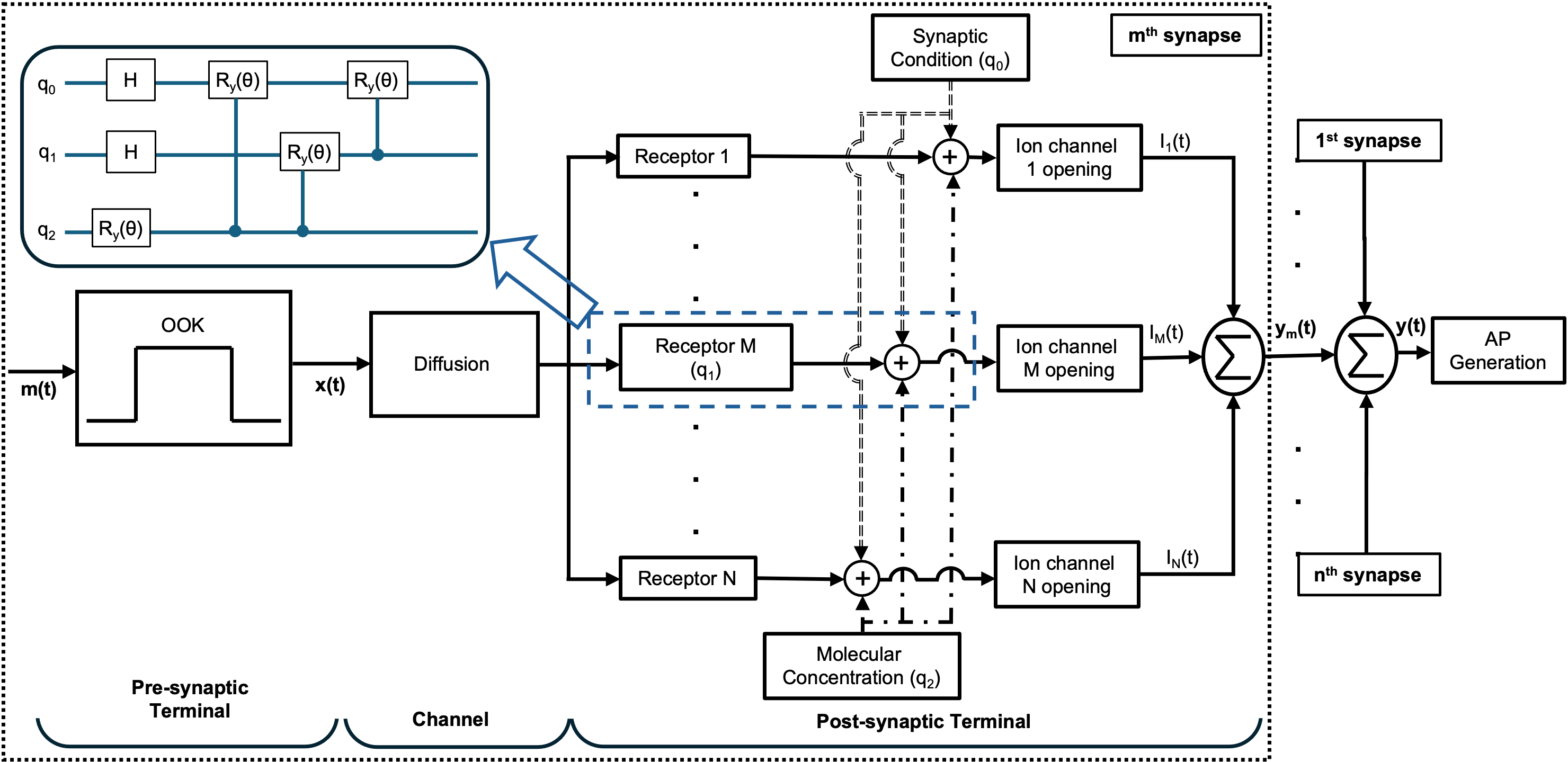}
    \caption{Block diagram of a Molecular Quantum (MolQ) Communication channel model.}
    \label{fig:GBAx_full}
\end{figure*}

Molecular Communication (MC) is a communication paradigm inspired by the cellular level \bluecite{gine2009molecular} biological communication that considers the Brownian Motion of molecules as a propagating mechanism rather than the conventional electromagnetic waves \bluecite{eckford2007nanoscale}. 
In MC, molecules known as Information Molecules (IM) are used to transceive information from the transmitter and the receiver by encoding, releasing into the medium, propagating, detecting, and decoding \bluecite{akyildiz2008nanonetworks}. 
Some widely studied IMs are proteins like insulin, nucleic acids (e.g., DNA and RNA), ions (e.g., Ca$^{2+}$, K$^+$), chemical messengers like Acetylcholine, volatile organic compounds (VOCs) (used in plant MC), pheromones, and nanoparticles \bluecite{kuscu2019transmitter}. 
Diffusion is the primary mode of transport for MC, and it is mainly classified as active transport (energy-dependent) and passive (energy-independent) diffusion. 
Drift is another mode of MC. 
There is a significant difference between these two: Diffusion is defined as Random movement that can move in any direction in spatial-temporal dynamics, whereas drift is unidirectional active transport \bluecite{malak2012molecular}.
In the biological context, simple, i.e., passive diffusion is often associated with active transport like molecular reuptake \bluecite{khan2017diffusion,koca2024modelling}, enzymatic degradation, and diffusing away.
In this paper, we consider passive diffusion and diffusion with active transport both exist in the synaptic cleft, and they are elaborated comparatively.

Some existing theories can explain neuronal function at different levels of complexity.
The simplest model is \textit{Integrate-and-fire model}, ignoring the biophysical mechanism \bluecite{dayan2005theoretical}.
One widely considered and mostly accepted model is \textit{Hodgkin-Huxley model} (H-H model), which explains the action potential (AP) generated by a neuron by considering the channel conductance \bluecite{hodgkin1952measurement,hodgkin1952currents,hodgkin1952components,loligo1952dual,hodgkin1952quantitative}.
Even though these models can explain the electrochemical mechanisms of a neuron, they do not consider neurotransmitter release, diffusion, and receptor-ligand binding.
Different researchers have examined the integration of these processes, specifically focusing on the physical channel model of neuro-spike communication \bluecite{veletic2016peer, balevi2013physical}.
The theoretical study of synaptic channel models is given in \bluecite{malak2013communication} and \bluecite{manwani2001detecting}.
Even though all the theories can justify the synaptic communication mechanism, those individual theories have significant limitations. 
The theoretical analysis of synaptic communication, considering MC, terminates at the POSM. 
Moreover, it can not explain the mechanism of ion channel formation. 
On the other hand, the H-H model is deterministic and does not consider the stochastic nature of synapse \bluecite{strassberg1993limitations}.  
Quantum Communication (QC) can give more insight into neuronal communication by mitigating the limitations of conventional theories and providing justification for ion channels opening and initiating communication waves inside neurons.

QC is a paradigm that is used to transfer a quantum state from one place to another. 
Quantum states encode quantum information, named qubits, represented in a two-dimensional Hilbert space. 
The quantum information that is transferred with the two fundamental principles of QC, superposition and entanglement, allows one to perform a function effortlessly, which is very complex in the classical sense \bluecite{gisin2007quantum}.
Explanation of neuronal communication in the non-classical communication paradigm is emerging as classical approaches failed to explain the brain's behavior.
Considering NMR techniques, Zero Quantum Coherence (ZQC) of the brain is explored in \bluecite{kerskens2022experimental}, and non-local ZQC is obtained, justifying the existence of quantum entanglement.
Furthermore, in \bluecite{liu2004local}, it has been experimented and validated that in cultured rat hippocampal neurons, the excitatory and inhibitory synaptic inputs can balance each other in a coordinated way if they are colocalized on the same dendritic branch. 
Inhibitory synapses can detect the impact of the excitatory synaptic inputs, which are colocalized on the same dendritic branch and determine whether to get activated or remain inhibited. 
This is, indeed, an experimental validation of the existence of the quantum entanglement at the post-synaptic membrane.
Moreover, there is some speculation on the quantum communication process at the neuronal synapses in \bluecite{xin2023decision, xin2023computational}, which indeed explores the quantum algorithm at the POSM, which is based on quantum decision making, built on quantum gates.
Even though the functionality of the elemental component of the brain, i.e., a neuron is known classically, the classical explanation fails to justify the main features of the brain, i.e., consciousness, memory, sleep, dreams, decision-making, etc.
There are some ground-breaking theories inspired by the quantum mechanism \bluecite{penrose2000large, stapp2004mind, fisher2015quantum, schwartz2005quantum, kerskens2022experimental, khrennikov2018quantum}, which have tried to explain the brain as a quantum communication system to validate or justify its functionality.
One widely accepted theory for consciousness is the Orchestrated Objective Reduction (Orch OR) theory, which suggests that human consciousness is carried out by the collection of microtubules inside the brain depending on the `orchestrated' quantum process. 
It also suggests that the OR is influenced by spacetime geometry, connecting the consciousness to the structure of the universe \bluecite{hameroff2014consciousness}. 
Hence, there is a requirement for the paradigm shift towards quantum communication to explain and understand the brain's behavior, memory, and consciousness \bluecite{koch2006quantum, georgiev2020quantum, georgiev2018quantum, kariev2021quantum}.

Thus, in this paper, we consider that after the release of neurotransmitters (acetylcholine) from the gut membrane into the synaptic cleft of GBAx, they diffuse through the cleft by MC and reach the VNM. 
For the analysis purpose, we have considered acetylcholine; however, this model is valid for all diffusion based signaling molecules and neurotransmitters.
Reaching the VNM causes the binding with the receptors and forms a ligand-receptor complex. 
These complexes lead to the opening of the ion channel of that neuron and initiate a communication wave. 
The whole process after molecules reach the VNM is described through the QC. 
Till now, there has been a significant research gap between the forming of ligand-receptor complex and the opening of ion channels. 
Ligand-receptor kinetics can be described by Michaelis–Menten kinetics \bluecite{michaelis1913kinetik}, and the H-H model can describe the AP initiation and propagation \bluecite{catterall2012hodgkin}, but there is no model that can explain the highly regulated opening and closing of ion channels. 
Here, we attempt to explain the highly coordinated and regulated channel opening through QC considering superposition and entanglement.
In summary, we consider a synapse of GBAx and propose a novel theoretical framework.
Molecular Quantum (MolQ) communication can explain the whole process, from releasing molecules from the gut membrane to opening the ion channels on the VNM.
Block diagram of MolQ communication model of Gut-Brain axis synapse is illustrated in Fig. \ref{fig:GBAx_full}.

The rest of the paper is organized as follows:
In Sec. \ref{sec:two}, details of physiological assumptions, the MC model to reach the VNM, and the formulation of a QC circuit for the opening of the ion channels after ligand-receptor binding on the VNM are given. 
A theoretical communication analysis of the proposed model is performed and discussed in Sec. \ref{sec:three}. 
The simulated results are described in Sec. \ref{sec:four}
Finally, we conclude this paper in Sec. \ref{sec:five}.

\section{GBAx Communication Channel and Ion Channel Opening}
\label{sec:two}
In this section, we propose our model for the transfer of information from the gut lumen to the brain through the 10$^{th}$ cranial nerve, i.e., the vagus nerve. 
We first describe the mathematical model of the MC associated with the transmission of molecules from the gut membrane to the VNM, followed by the hypothesis and relevant model development by considering QC on the VNM. 
A quantum circuit is proposed that can efficiently describe the complexity involved in the opening of the ion channel and the initiation of neuronal communication.

\subsection{Vesicle Release and Diffusion}
\label{sec:Vesicle}
The small intestine is covered with villi, and each villus is covered with a single layer of epithelium. 
This layer comprises different cells, and enteroendocrine (EEC) cells are one of them, often called the gut sensor; among these EEC cells, some cells synapse with the vagus nerve, called neuropod cells. 
These neuropod cells are responsible for sending the gut information to the brain. 
After sensing any movement or due to any hormonal activities, these neuropod cells can convert that sensory information into electrical impulses. 
After the arrival of electrical impulses, i.e., the AP, the vesicles are fused with the gut membrane \bluecite{kaelberer2020neuropod}.
In response, it performs vesicle-wise release of neurotransmitters like glutamate, GABA \bluecite{kaelberer2020neuropod}, and acetylcholine \bluecite{morkl2023gut}.
The number of vesicles that are released into the synaptic cleft widely varies depending upon the shape of the electrical impulses.
It can be considered as the On-Off Keying (OOK) modulation technique described in \bluecite{atakan2010molecular, atakan2007information}, where information molecules (IMs) are released into the environment when the transmitter is ``on," i.e., binary `1', and no IM is released while it is ``off."
In this model, we consider the central synapse and only one ``on" period of OOK, where the vesicle is released from the center of the gut membrane, and for the simplification of the model, we further consider that only one vesicle is released at a time.

Released neurotransmitters can diffuse through the synaptic cleft with a Brownian Motion.
It makes a concentration variation $C(R,t)$ across a three-dimensional space and time, which can be well approximated by Fick's second law of diffusion:
\begin{align}
    \frac{\partial C(R,t)}{\partial t} = D \times \nabla^2 C(R,t),
    \label{fick}
\end{align}
where $D$ is the diffusion coefficient, $R$ is a point in 3-D coordinate system, and $\nabla^2$ is the Laplace operator in 3-D Cartesian coordinates. 
\textcolor{blue}{(\ref{fick})} can be solved by considering suitable boundary and initial conditions. 
Here, we are considering a 3-D unbounded environment with an impulse release point, a widely studied case in MC \bluecite{kilinc2013receiver}. 
The initial condition of this PDE can be expressed as follows:
\begin{align}
    C(R_0, t = t_0) = N_0 \delta(R-R_0),
    \label{impulse}
\end{align}
where diffusion initiates with an instantaneous release of $N_0$ neurotransmitters from the gut membrane, i.e., $R_0$ at time $t_0$ and propagates to the VNM ($R$), and $\delta(\cdot)$ is the dirac delta function. 
The boundary condition can be said that there will be no flux of the neurotransmitters through the gut membrane or VNM
\begin{align}
\lim_{R \to 0} \frac{\partial C(R,t)}{\partial t} = \lim_{R \to \infty} \frac{\partial C(R,t)}{\partial t} = 0,
\label{boundary}
\end{align}
where $R$$\to$$0$ signifies the gut membrane, and $R$$\to$$\infty$ implies the VNM.
Solving \textcolor{blue}{(\ref{fick})} with the help of the initial and the boundary conditions that are given in \textblue{(\ref{impulse})} and \textblue{(\ref{boundary})}, the obtained equation is as follows:
\begin{align}
    C(R,t)= \frac{N_0}{({4\pi D T})^{\frac{3}{2}}} e^{-\frac{{\|R - R_0\|}^2}{4DT} },
    \label{solved}
\end{align}
where $T$ is the time taken by the neurotransmitters to reach the VNM from the gut membrane.

Although the active diffusion is less studied, a model of active transport considering molecular reuptake exists in \bluecite{khan2017diffusion} and is given by,
\begin{align}
&C_{ru}(R,t) = \frac{N_0}{{(4\pi D t)}^{3/2}} e^{\frac{-x^2-y^2}{4Dt}} \times  \left\{ \sum_{k=-\infty}^{-1}  (2-P_{ru}) \right. \nonumber &\\
& \left. (1-P_{ru})^{-k-1} e^\zeta  + \sum_{k=0}^{\infty}  (2-P_{ru})  (1-P_{ru})^{k} e^\zeta \right\},
\label{reuptake}
\end{align}
where $\zeta={\frac{-(z-(2k+1)h)^2}{4Dt}}$, $h$ is the synaptic cleft height, $P_{ru}$ is the probability of reuptake. 
Moreover, we can incorporate diffusion away from the cleft as advection, which is given by $v \nabla C(R,t)$, where $v$ is the advection velocity. 
Incorporation of diffusion away keeps \textcolor{blue}{(\ref{reuptake})} unchanged, affecting $\zeta$ as $\zeta = {\frac{-(z-(2k+1)h-vt)^2}{4Dt}}$.

\subsection{Initiation of Ion Channel}
\label{sec:ion}
After the neurotransmitters reach the VNM, the following process of the ion channel initiation starts. 
In this section, we will propose a model for the ion channel initiation. 

Neural synapse is dynamic, an interplay of the excitatory and inhibitory synaptic conditions \bluecite{liu2004local}. The synapse excites or inhibits depending on the ligand-receptor complex \bluecite{niyonambaza2019review} and the previously produced action potential \bluecite{glasgow2019approaches}. 
Hence, the synapse is a superposition of two states, i.e., excitatory and inhibitory. 
So, the superposed state of the synapse can be expressed as:
\begin{align}
    |\psi\rangle_0 = c_{0_0} |q_{0_0}\rangle + c_{0_1} |q_{0_1}\rangle,
\end{align}
where $|\psi\rangle_0$ is the state of a synapse.
$|q_{0_0}\rangle$ and $|q_{0_1}\rangle$ denote the excited and inhibited state of the synapse, respectively, and $|c_{0_0|^2}$ and $|c_{0_1}|^2$ are probabilities of excitation and inhibition of synapse, respectively, where $|c_{0_0}|^2$$+$$|c_{0_1}|^2$$=$$ 1$.
Furthermore, when a ligand binds with a receptor, it can either excite the receptor to create an ion channel or remain inhibited depending on the neurotransmitter concentration or the formation of a ligand-receptor complex.
The superposed state of the receptor can be expressed as:
\begin{align}
    |\psi\rangle_1 = c_{1_0} |q_{1_0}\rangle + c_{1_1} |q_{1_1}\rangle,
\end{align}
where $|\psi\rangle_1$ is the state of the receptor.
$|q_{1_0}\rangle$ and $|q_{1_1}\rangle$ denote the excite and inhibit state of the receptor, respectively, and driven by the probability of excitation and inhibition, given by $|c_{1_0}|^2$ and $|c_{1_1}|^2$, respectively, where $|c_{1_0}|^2$$+$$|c_{1_1}|^2$$=$$ 1$.

In this design, we will consider an equal superposed state, which can easily be produced by Hadamard ($H$) quantum gate. 
It simplifies the initial state preparation and allows the model to generalize the behavior of synaptic and receptor states without bias towards excitation or inhibition.
However, the assumption of equal superposition may not always hold in biological systems, where initial conditions can vary due to prior synaptic responses.
$H$ gate is defined by the following matrix:
\begin{equation}
\label{eqn:matrix}
H
=
\frac{1}{\sqrt{2}}
\begin{bmatrix}
1 & 1\\ 1 & -1
\end{bmatrix},
\end{equation} 
The output state of $H$ depends on the input qubit as follows:
\begin{align}
    H|0\rangle = \frac{1}{\sqrt{2}}(|0\rangle + |1\rangle), \label{eq:H0}\\
    H|1\rangle = \frac{1}{\sqrt{2}}(|0\rangle - |1\rangle)\label{eq:H1},
\end{align}
\textcolor{blue}{(\ref{eq:H0})} and \textcolor{blue}{(\ref{eq:H1})} show that the $H$ gate produces an equal superposition of the two orthonormal basis states.

Moreover, the information is sent from the gut membrane through $C(R,t)$ to the VNM. 
We will consider the concentration of neurotransmitter as an ancilla qubit ($|q_2\rangle$), which is not a direct part of the model, but it will indirectly control the output state, i.e., the opening of ion channels. 
$|q_2\rangle$ allows the quantum model to dynamically respond to changes in neurotransmitter levels, representing the biological process of synaptic communication and ion channel regulation.
The rotation of $|q_2\rangle$ is performed by incorporating a $R_y$ quantum gate; the superposition state ($|\psi\rangle_2$) will be as follows:
\begin{equation}
\label{eqn:matrix}
|\psi\rangle_2
=
R_y
\begin{bmatrix}
    |q_{2_0}\rangle \\ |q_{2_1}\rangle
\end{bmatrix}, \hspace{4pt}
R_y = \begin{bmatrix}
\cos\left(\frac{\theta}{2}\right) & -\sin\left(\frac{\theta}{2}\right) \\[4pt]
\sin\left(\frac{\theta}{2}\right) & \cos\left(\frac{\theta}{2}\right)
\end{bmatrix},
\end{equation} 
where $|q_{2_0}\rangle$ and $|q_{2_1}\rangle$ corresponds to high and low concentration profile respectively.
The angle of the rotation ($\theta$) is determined by the concentration of the neurotransmitters as follows,
\begin{align}
    \theta = \frac{C-C_{min}}{C_{max}-C_{min}} \cdot 2\pi,
\end{align}
where $C$ is a spatial-temporal neurotransmitter concentration, $C_{min}$ is the minimum concentration, $C_{max}$ is the maximum obtained concentration.

\begin{figure}
    \centering
    \begin{quantikz}
    \lstick{$q_0$} & \gate{H}     & \gate{R_y(\theta)}    &   \qw        & \gate{R_y(\theta)}   & \meter{} & \qw \\
    \lstick{$q_1$} & \gate{H}     &     \qw     & \gate{R_y(\theta)}     & \ctrl{-1} & \meter{} & \qw \\
    \lstick{$q_2$} & \gate{R_y(\theta)} & \ctrl{-2} & \ctrl{-1} &  \qw         & \meter{} & \qw \\
    \end{quantikz}
    \caption{Quantum communication circuit of the VNM ion channel initiation.}
    \label{fig:QC circuit}
\end{figure}
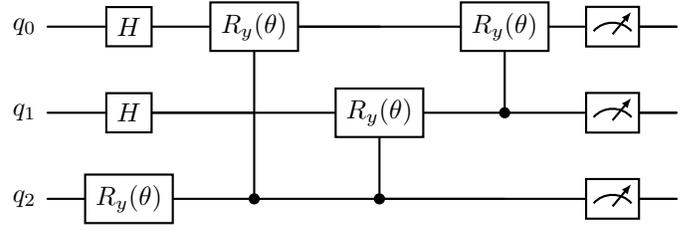

$|q_2\rangle$ directly influences the overall output. The incoming neurotransmitter concentration affects the synapse and the neurotransmitters also bind with receptors. Hence, $|q_2\rangle$ is entangled with $|q_0\rangle$ and $|q_1\rangle$, impacting their combined state. 
The entanglement between $|q_2\rangle$ and the primary qubits provides a mechanism to capture the inter-dependencies between neurotransmitter concentration and synaptic activity, reflecting the biological interactions between the pre-synaptic and post-synaptic membranes, thereby influencing the output state of the quantum model.
Furthermore, the formation of ligand-receptor complex also influences the synaptic condition. 
Hence, $|q_1\rangle$ is entangled with $|q_0\rangle$. 
The entanglement is produced using the Controlled-R\textsubscript{y} ($CR_y$) gate. 
This gate acts on the target qubit according to the state of the control qubit, where if the control qubit is $|0\rangle$; the target qubit will remain the same, but if the control qubit is $|1\rangle$, it will incorporate an amplitude and phase shift on the target qubit. 
$CR_y$ entangles the control qubit with the target qubit.
The resulting entangled state reflects the combined influence of the neurotransmitter concentration and the initial superposition states of the synapse and receptor, impacting the probability amplitudes of the qubit states.
Biologically, the probability of neural synapse excitation and receptor binding depends on available neurotransmitter concentration \cite{franks2003independent}.
Through $CR_y$ gates, we attempted to capture the dynamic influences of neurotransmitter concentration on synapse and receptors.
The output state of $CR_y$ is as follows:
\begin{align}
    CR_y|0\rangle = cos\frac{\theta}{2}|0\rangle + sin\frac{\theta}{2}|1\rangle, \label{eq:Ry0}\\
    CR_y|1\rangle = -sin\frac{\theta}{2}|0\rangle + cos\frac{\theta}{2}|1\rangle, \label{eq:Ry1}
\end{align}
The designed quantum communication circuit is depicted in Fig. \ref{fig:QC circuit}.

\section{Theoretical Analysis of Quantum Communication}
\label{sec:three}
During quantum circuit analysis, qubits are initialized with $|0\rangle$. 
Hence, considering all qubits are initialized with $|0\rangle$, the combined state of $|q_0\rangle$ and $|q_1\rangle$ after $H$ can be expressed by the tensor product of their individual states can be given by:
\begin{align}
    H|q_0\rangle \otimes H|q_1\rangle &= \left(\frac{1}{\sqrt{2}}(|0\rangle + |1\rangle) \right) \otimes \left(\frac{1}{\sqrt{2}}(|0\rangle + |1\rangle) \right) \nonumber \\
    &= \frac{1}{2} (|00\rangle + |01\rangle +|10\rangle + |11\rangle)   
\end{align}
Considering $|q_3\rangle$ after the application of $R_y$, the combined state becomes:
\begin{align}
    &\frac{1}{2} (|00\rangle + |01\rangle +|10\rangle + |11\rangle) \otimes (P |0\rangle + Q |1\rangle)) \nonumber \\
    & \quad \quad = \frac{1}{2} (P(|000\rangle + |010\rangle + |100\rangle +  |110\rangle) \nonumber \\ 
    &\quad \quad \quad + (Q(|001\rangle + |011\rangle + |101\rangle +  |111\rangle))
\end{align}
where, $P=cos\frac{\theta}{2}$ and $Q=sin\frac{\theta}{2}$. In the circuit, there are three CR\textsubscript{y} gates; first, $|q_0\rangle$ is controlled by $|q_2\rangle$; next, $|q_1\rangle$ is controlled by $|q_2\rangle$, and the last one is from $|q_1\rangle$ to $|q_0\rangle$.
The output state after the first CR\textsubscript{y} gate can be given by:
\begin{align}
    |\psi\rangle_{C_{1}} &= \frac{1}{2} [P(|000\rangle + |010\rangle +|100\rangle + |110\rangle) + Q(P-Q)\nonumber \\
    & (|001\rangle + |011\rangle) + Q(P+Q) (|101\rangle + |111\rangle)],
\end{align}
This state will act as the input state of the second controlled R\textsubscript{y} gate. The output state of the second gate $|\psi\rangle_{C_{2}}$ can be expressed as follows:
\begin{align}
    |\psi\rangle_{C_{2}} =& \frac{1}{2} [P(|000\rangle + |010\rangle +|100\rangle + |110\rangle) + Q(P-Q)\nonumber \\
    &((P-Q)|001\rangle + (P+Q)|011\rangle) +  
    Q(P+Q) \nonumber \\ &((P-Q)|101\rangle + (P+Q)|111\rangle)],
\end{align}
The final output state of the QC can be expressed as follows:
\begin{align}
    |\psi\rangle_{C_{f}} =& \frac{1}{2} [P|000\rangle - Q(P-Q)^2|001\rangle + P(P-Q) |010\rangle \nonumber \\
     & \negmedspace + (QP(P^2-3Q^2)-Q^2(P^2+Q^2)) |011\rangle + P |100\rangle \nonumber \\ 
     & \quad \quad + Q(P^2-Q^2)|101\rangle + P(P+Q) |110\rangle + \nonumber \\ 
     & \quad (PQ(P^2+Q^2)+Q^2(3P^2-Q^2)) |111\rangle)],
\end{align}
In our study, we will specifically use the $|111\rangle$ state to claim the successful opening of the ion channel. However, the conductance of an ion channel depends upon the number of ions that pass through the channel. 
The conductance can be characterized by the number of successful shots of $|111\rangle$ for a given concentration; the higher the number of shots signifies, the higher the conductance of the channel. 
In a quantum circuit simulation, shots refer to the number of times a quantum circuit is executed to collect measurement results.

Moreover, the designed QC model is discussed in terms of information technology and we have analyzed the entropy and the mutual information to validate the communication channel.
The entropy ($H(|111\rangle)$) of the system can be expressed as 
\begin{align}
    H(|111\rangle) = - (p(|111\rangle) &\log_2 p(|111\rangle) \nonumber \\ &+ p((|111\rangle)^c) \log_2 p(|111\rangle)^c),
    \label{eq:en}
\end{align}
where $p(|111\rangle)$ is the probability of the $|111\rangle$, which signifies the opening of the ion channel; and $p((|111\rangle)^c)$ indicates probability of the channel remaining closed. 
In this design, we consider that only two outcomes are possible.

Furthermore, Mutual Information (MI), which quantifies the reduction in uncertainty about one variable ($|111\rangle$) given knowledge of the other ($C(R,t)$), is determined. 
The MI ($I(C;|111\rangle)$) can be expressed as follows:
\begin{align}
    I(C;|111\rangle) = \sum_{c \in C} \sum_{l \in |111\rangle} p(c, l) \log_2 \left( \frac{p(c, l)}{p(c) p(l)} \right),
    \label{eq:mi_g}
\end{align}
where $p(c, l)$ is the joint probability distribution function of $C(R,t)$ and $|111\rangle$, and $p(c)$ and $p(l)$ are the marginal probability distribution functions of $C(R,t)$ and $|111\rangle$ respectively.

\section{Results and Discussion}
\label{sec:four}
In this section, we demonstrate the output state of the ion channel, considering the concentration of the neurotransmitters as the controller. 
Furthermore, we validate the quantum communication model by discussing the entropy and the mutual information transfer.

\begin{table}[t!]
    \centering
    \caption{Simulation parameters}
    \begin{tabular}{lcl}
    \hline
       \textit{Neurotransmitters in a vesicle}  & $N_0$ & 10000 \\
        \textit{Synaptic cleft height} & $-$ & 50nm\\
        \textit{VNM area} & L$\times$W & 500nm $\times$ 500nm\\
        \textit{Diffusion coefficient} & $D$ & 4$\times$10$^{-10}$ m$^{2}$/s \bluecite{land1984diffusion}\\
        \textit{Total simulation time} & $-$ & 500$\mu$s \\
        \textit{Synaptic reuptake} & $P_{ru}$ & 10\% \bluecite{khan2017diffusion}\\
        \textit{Advection velocity} & $v$ & 1$\mu$m/s \bluecite{wei2022changes}\\
        \textit{Number of shots} & $-$ & 1000\\
     \hline
    \end{tabular}
    \label{tab:para}
\end{table}

\begin{figure}[t!]
\centering
\begin{subfigure}[b]{0.15\textwidth}
    \includegraphics[width=\textwidth]{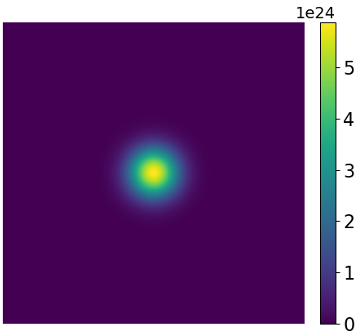} 
    \caption{}
\end{subfigure}
\hfill
\begin{subfigure}[b]{0.15\textwidth}
    \includegraphics[width=\textwidth]{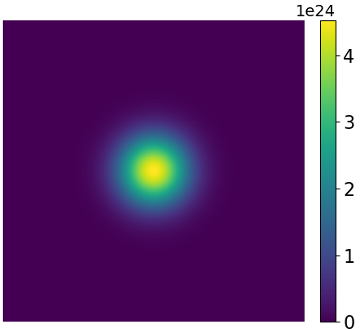} 
    \caption{}
\end{subfigure}
\hfill
\begin{subfigure}[b]{0.15\textwidth}
    \includegraphics[width=\textwidth]{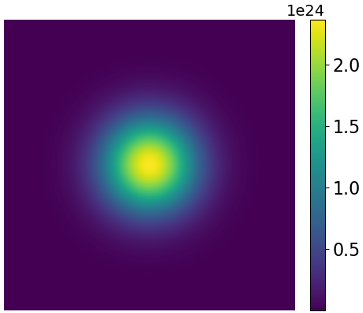} 
    \caption{}
\end{subfigure}

\vspace{0.5cm} 

\begin{subfigure}[b]{0.15\textwidth}
    \includegraphics[width=\textwidth]{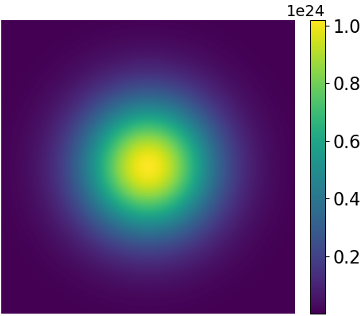} 
    \caption{}
\end{subfigure}
\hfill
\begin{subfigure}[b]{0.15\textwidth}
    \includegraphics[width=\textwidth]{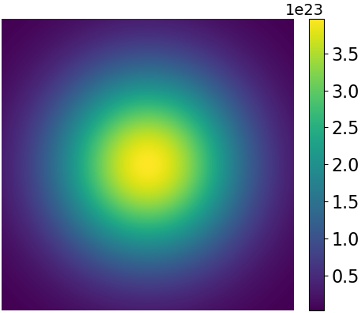} 
    \caption{}
\end{subfigure}
\hfill
\begin{subfigure}[b]{0.15\textwidth}
    \includegraphics[width=\textwidth]{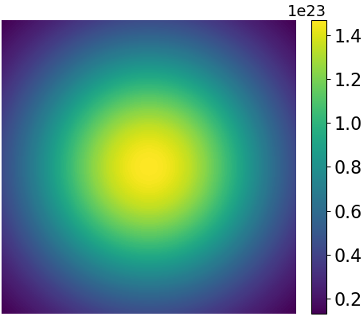} 
    \caption{}
\end{subfigure}

\vspace{0.5cm} 

\begin{subfigure}[b]{0.15\textwidth}
    \includegraphics[width=\textwidth]{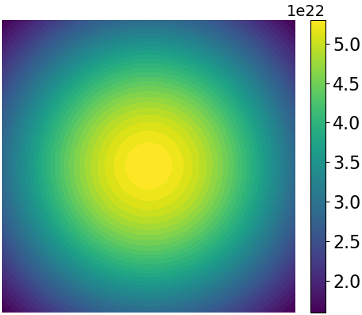} 
    \caption{}
\end{subfigure}
\hfill
\begin{subfigure}[b]{0.15\textwidth}
    \includegraphics[width=\textwidth]{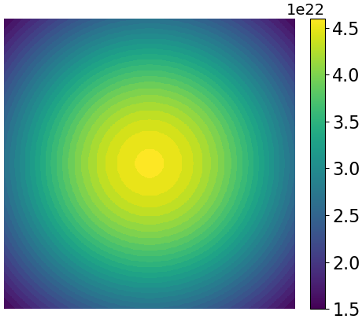} 
    \caption{}
\end{subfigure}
\hfill
\begin{subfigure}[b]{0.15\textwidth}
    \includegraphics[width=\textwidth]{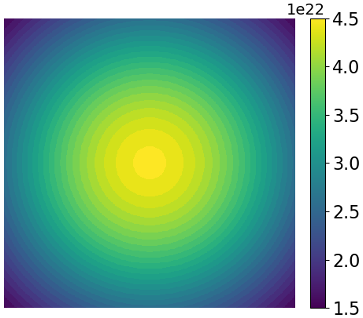} 
    \caption{}
\end{subfigure}

\caption{Time-course of neurotransmitter concentration (molecules/m$^2$) at the VNM at (a) 1 $\mu$s, (b) 2 $\mu$s, (c) 4 $\mu$s, (d) 8 $\mu$s, (e) 16 $\mu$s, (f) 32 $\mu$s, (g) 64 $\mu$s, (h) 71 $\mu$s, (i) 72 $\mu$s.}
\label{fig:conc}
\end{figure}

\begin{figure}[t!]
\centering
\begin{subfigure}[b]{0.15\textwidth}
    \includegraphics[width=\textwidth]{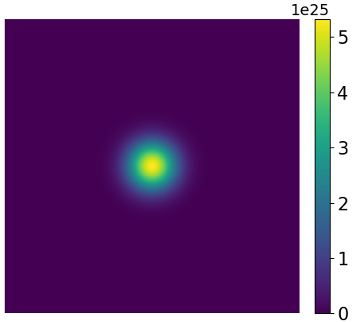} 
    \caption{}
\end{subfigure}
\hfill
\begin{subfigure}[b]{0.15\textwidth}
    \includegraphics[width=\textwidth]{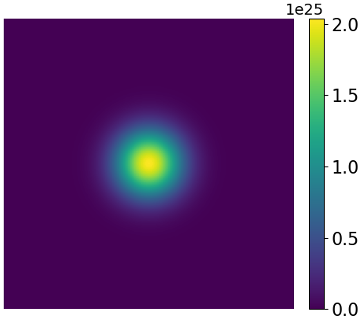} 
    \caption{}
\end{subfigure}
\hfill
\begin{subfigure}[b]{0.15\textwidth}
    \includegraphics[width=\textwidth]{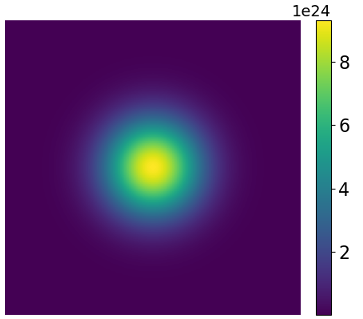} 
    \caption{}
\end{subfigure}

\vspace{0.5cm} 

\begin{subfigure}[b]{0.15\textwidth}
    \includegraphics[width=\textwidth]{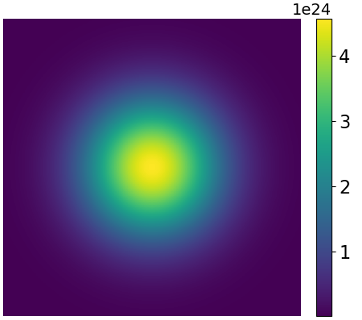} 
    \caption{}
\end{subfigure}
\hfill
\begin{subfigure}[b]{0.15\textwidth}
    \includegraphics[width=\textwidth]{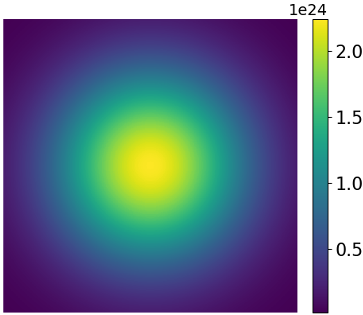} 
    \caption{}
\end{subfigure}
\hfill
\begin{subfigure}[b]{0.15\textwidth}
    \includegraphics[width=\textwidth]{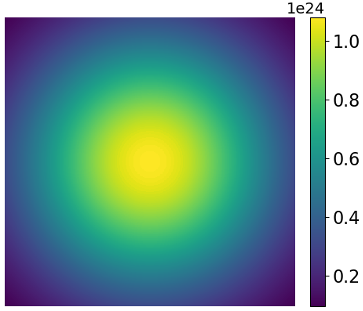} 
    \caption{}
\end{subfigure}

\caption{Time-course of active diffusion based neurotransmitter concentration (molecules/m$^2$) at the VNM at (a) 1 $\mu$s, (b) 2 $\mu$s, (c) 4 $\mu$s, (d) 8 $\mu$s, (e) 16 $\mu$s, (f) 32 $\mu$s.}
\label{fig:active_c}
\end{figure}

In Sec. \ref{sec:Vesicle}, the molecular communication mechanism considering passive diffusion to reach the VNM is discussed, and the simulated results can be seen in Fig. \ref{fig:conc}, and  Fig. \ref{fig:active_c} illustrates the communication mechanism of diffusion associated with active transport.
The parameters considered for the simulation are tabulated in Table \ref{tab:para}. 
The maximum concentration occurs at 1 $\mu$s and slowly decays after that. 
The results clearly depict that the higher concentration can be obtained at the center of the VNM, exactly perpendicular to the point from where the neurotransmitters are released.
This concentration is considered as the input of the QC model. 
However, if we consider the spatial-temporal distribution, it is an array of 500$\times$500$\times$500. 
The whole model becomes computationally intensive. 
For this reason, we have defined a threshold of 1$\times$10$^{21}$ molecules/m$^2$. 
Above the threshold, the multiple decimal points are approximated to two, resulting in a reduction in the handling of large data.

\begin{figure}[t!]
\centering
\begin{subfigure}[b]{0.15\textwidth}
    \includegraphics[width=\textwidth]{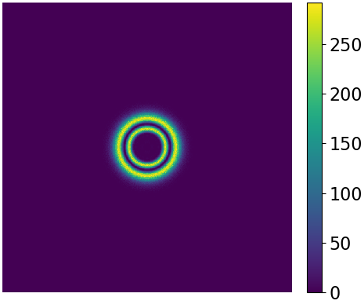} 
    \caption{}
\end{subfigure}
\hfill
\begin{subfigure}[b]{0.15\textwidth}
    \includegraphics[width=\textwidth]{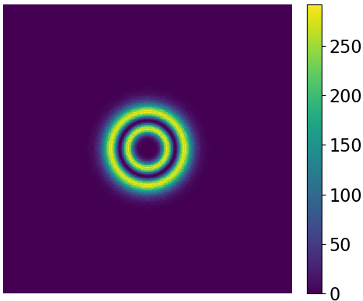} 
    \caption{}
\end{subfigure}
\hfill
\begin{subfigure}[b]{0.15\textwidth}
    \includegraphics[width=\textwidth]{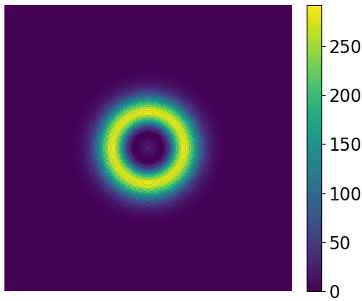} 
    \caption{}
\end{subfigure}

\vspace{0.5cm} 

\begin{subfigure}[b]{0.15\textwidth}
    \includegraphics[width=\textwidth]{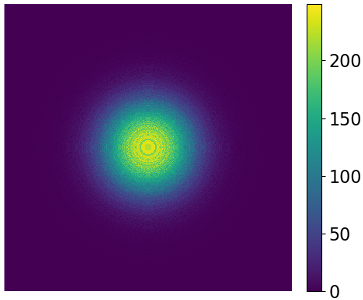} 
    \caption{}
\end{subfigure}
\hfill
\begin{subfigure}[b]{0.15\textwidth}
    \includegraphics[width=\textwidth]{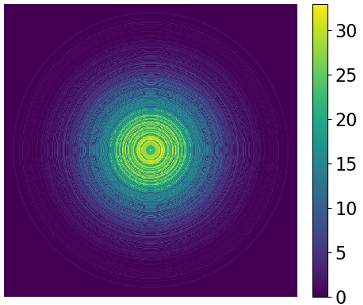} 
    \caption{}
\end{subfigure}
\hfill
\begin{subfigure}[b]{0.15\textwidth}
    \includegraphics[width=\textwidth]{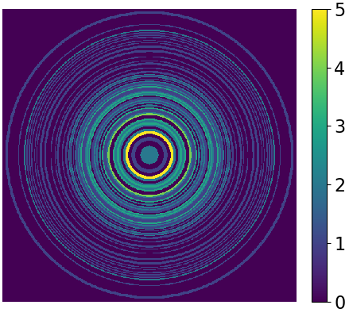} 
    \caption{}
\end{subfigure}

\vspace{0.5cm} 

\begin{subfigure}[b]{0.15\textwidth}
    \includegraphics[width=\textwidth]{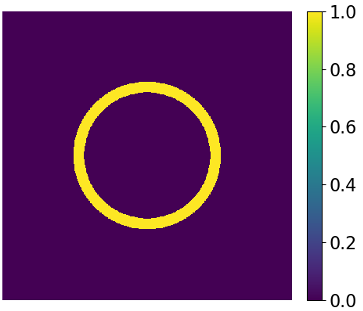} 
    \caption{}
\end{subfigure}
\hfill
\begin{subfigure}[b]{0.15\textwidth}
    \includegraphics[width=\textwidth]{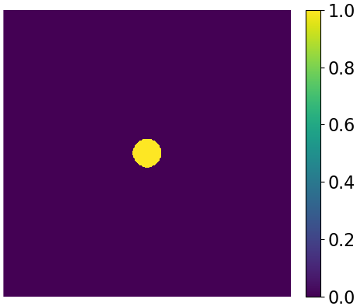} 
    \caption{}
\end{subfigure}
\hfill
\begin{subfigure}[b]{0.15\textwidth}
    \includegraphics[width=\textwidth]{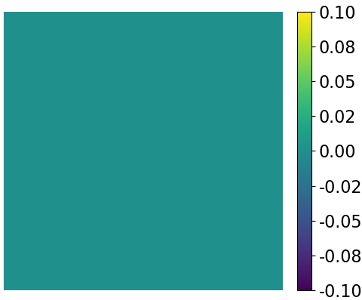} 
    \caption{}
\end{subfigure}

\caption{The outcome of the quantum communication model with 1000 shots at (a) 1 $\mu$s, (b) 2 $\mu$s, (c) 4 $\mu$s, (d) 8 $\mu$s, (e) 16 $\mu$s, (f) 32 $\mu$s, (g) 64 $\mu$s, (h) 71 $\mu$s, (i) 72 $\mu$s.}
\label{fig:count}
\end{figure}

\begin{figure}[t!]
\centering
\begin{subfigure}[b]{0.15\textwidth}
    \includegraphics[width=\textwidth]{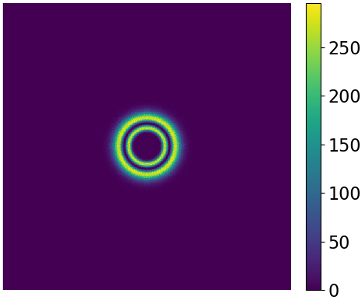} 
    \caption{}
\end{subfigure}
\hfill
\begin{subfigure}[b]{0.15\textwidth}
    \includegraphics[width=\textwidth]{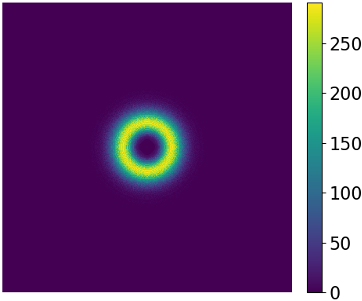} 
    \caption{}
\end{subfigure}
\hfill
\begin{subfigure}[b]{0.15\textwidth}
    \includegraphics[width=\textwidth]{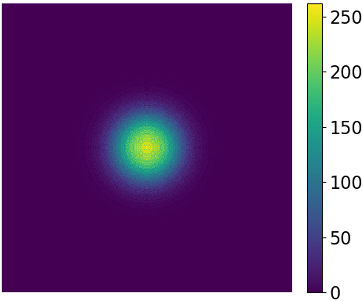} 
    \caption{}
\end{subfigure}

\vspace{0.5cm} 

\begin{subfigure}[b]{0.15\textwidth}
    \includegraphics[width=\textwidth]{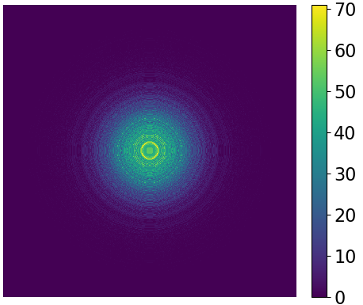} 
    \caption{}
\end{subfigure}
\hfill
\begin{subfigure}[b]{0.15\textwidth}
    \includegraphics[width=\textwidth]{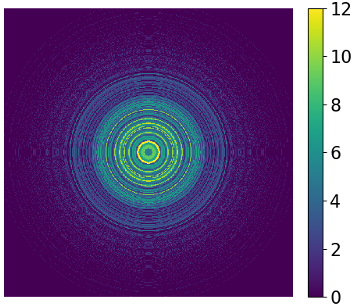} 
    \caption{}
\end{subfigure}
\hfill
\begin{subfigure}[b]{0.15\textwidth}
    \includegraphics[width=\textwidth]{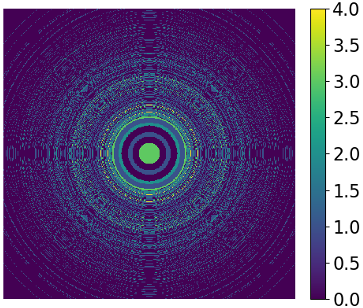} 
    \caption{}
\end{subfigure}

\caption{The outcome of the quantum communication model considering active diffusion at (a) 1 $\mu$s, (b) 2 $\mu$s, (c) 4 $\mu$s, (d) 8 $\mu$s, (e) 16 $\mu$s, (f) 32 $\mu$s.}
\label{fig:active_q}
\end{figure}

\begin{figure}[t!]
    \centering
    \includegraphics[width=0.4\textwidth]{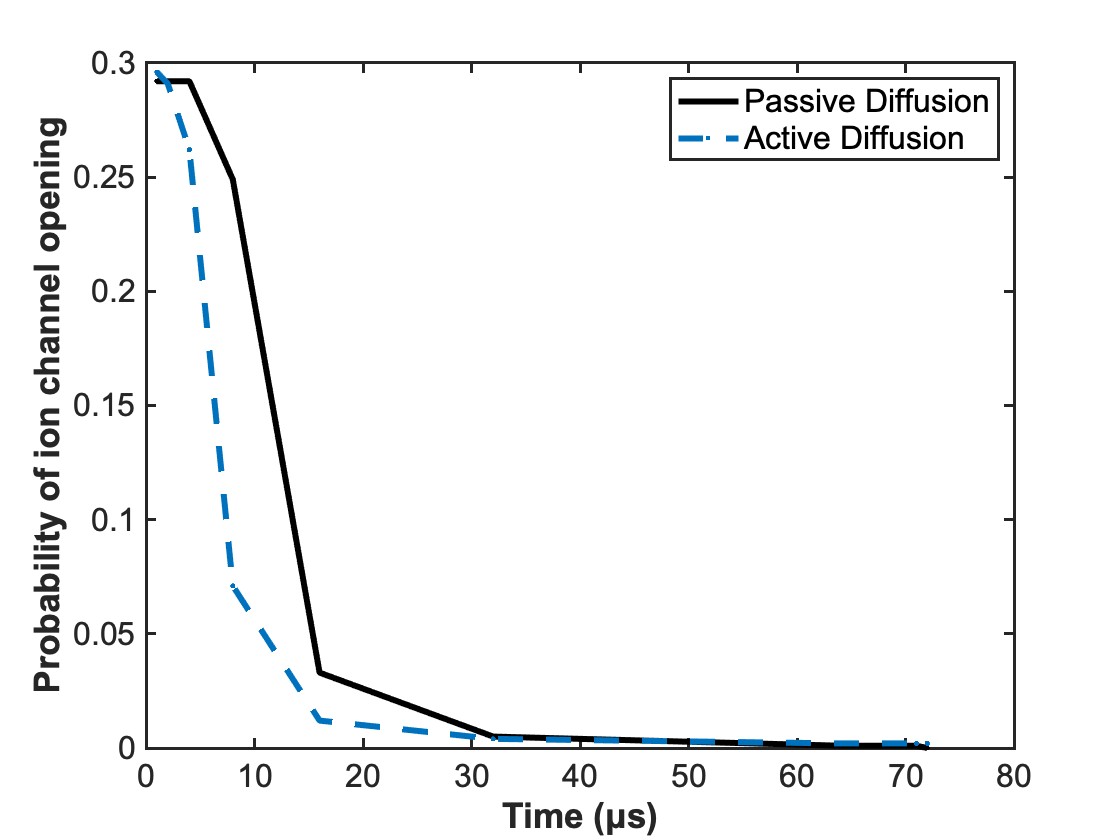}
    \caption{Variation of ion channel opening probability at the VNM.}
    \label{fig:prob}
\end{figure}

Next, we have simulated the quantum circuit mentioned in Sec. \ref{sec:ion}. 
We are considering the $|111\rangle$ state as the successful opening of an ion channel. 
By simulating the QC model with the parameter mentioned in Table \ref{tab:para}, the obtained results are depicted in Fig. \ref{fig:count} and Fig. \ref{fig:active_q}, considering passive diffusion and active transport, respectively.
These results suggest that just after the release of the neurotransmitters from the gut membrane, the VNM creates an ion channel at the center for information transmission. 
The ion channel opening slowly spreads around, as shown in Fig. \ref{fig:count}a,b,c and Fig. \ref{fig:active_q}a,b,c.
The ion channel steadily closes over time, and ultimately, after 71 $\mu$s, there is no opened ion channel in the passive diffusion based model, as shown in Fig. \ref{fig:count}i. 
It is also interesting to note that the probability of ion channel opening is the highest initially and rapidly decreases over time, as shown in Fig. \ref{fig:prob}. 
These results can demonstrate the capability of fast initiation of action potential in the VN neuron.  

If we consider the concentration profile on the VNM at 256 $\mu$s, as depicted in Fig. \ref{fig:conc}i, we can clearly mention that there is still a high concentration of neurotransmitters present to bind with the receptor. 
However, Fig. \ref{fig:count} clearly mentions that after 71 $\mu$s, there is 0 probability of ion channels opening. 
Hence, the remaining neurotransmitters should be cleared from the synapse. 
This can be described by the termination mechanisms of molecular communication, like enzymatic degradation or diffusing away from the cleft.
In other words, the probability of the ion channel opening depicts that the information mainly gets transferred in the initial time frames after the firing.
The incorporation of active transport mechanisms in passive diffusion allows us to understand their individual and combined effects on neurotransmitter concentration. 
Our comparative study shows that the inclusion of active transport mechanisms does not significantly interfere with the results derived from passive diffusion alone but provides a more nuanced understanding of synaptic processes.
The primary effect of including active diffusion is a more accurate representation of neurotransmitter dynamics.

\begin{figure}[t!]
\centering
\begin{subfigure}[b]{0.225\textwidth}
    \includegraphics[width=\textwidth]{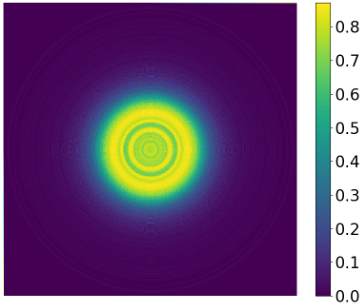} 
    \caption{}
\end{subfigure}
\hfill
\begin{subfigure}[b]{0.215\textwidth}
    \includegraphics[width=\textwidth]{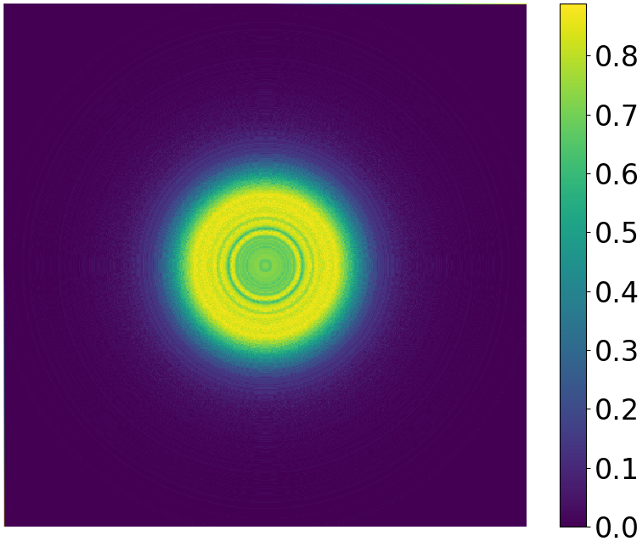} 
    \caption{}
\end{subfigure}
\caption{Maximum temporal entropy (bits) distribution on VNM considering (a) Passive diffusion, and (b) Diffusion with active transport.}
\label{fig:entropy}
\end{figure}

The designed quantum communication circuit is analyzed in terms of information technology, where the entropy is determined. 
We consider the output of the QC to have two outcomes: one is open, which is characterized by the $|111\rangle$ state, and all other states define the closed state of the ion channels. 
Using \textblue{(\ref{eq:en})}, the entropy is measured, and if all the states are equally likely, the theoretical entropy becomes 0.544 bits. 
However, the maximum entropy achieved by the proposed model for passive diffusion is 0.871 bits, as shown in Fig. \ref{fig:entropy}a, and considering passive diffusion with active transport is 0.8885 bits, as illustrated in Fig. \ref{fig:entropy}b.
Obtained results signify that the information that is being transmitted from the gut membrane to VNM is significantly high. 
High entropy in active transport, in comparison with passive diffusion, indicates that the active diffusion process allows for a broader range of states and uncertainties compared to passive diffusion.
It can imply a robust capacity for the synapse to encode and transmit a wide range of information about gut conditions to the brain.
High entropy indicates the system's capacity to adapt to a wide range of signals from the gut, essential for responding to varying conditions such as changes in gut microbiota composition.

\begin{figure}[t!]
\centering
\begin{subfigure}[b]{0.22\textwidth}
    \includegraphics[width=\textwidth]{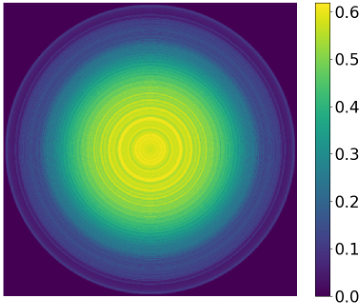} 
    \caption{}
\end{subfigure}
\hfill
\begin{subfigure}[b]{0.22\textwidth}
    \includegraphics[width=\textwidth]{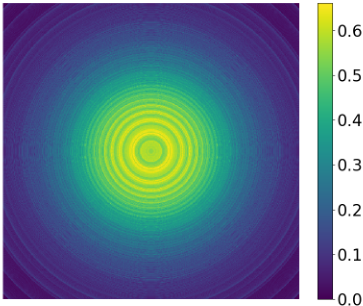} 
    \caption{}
\end{subfigure}
\caption{Spatial distribution of temporal mutual information (bits) on VNM (a) Passive diffusion, and (b) Diffusion with active transport.}
\label{fig:MI}
\end{figure}

Furthermore, MI of the concentration and the ion channel opening for the initiation of communication in the VN neuron is studied. 
Using \textblue{(\ref{eq:mi_g})}, the obtained MI for passive diffusion is depicted in Fig. \ref{fig:MI}a, which has the maximum obtained MI is 0.6182 bits. 
It indicates that around 71\% of the ion channel state can be explained by the input signal. 
MI for active transport associated with passive diffusion is depicted in Fig. \ref{fig:MI}b.
The maximum obtained MI is 0.6609 bits. 
It depicts approximately 74\% of the model output state can be inferred from the input signal.
This strong relationship suggests effective communication and control within the neural system.
The enhanced MI for active diffusion can indicate better coordination and regulation within the system, as active transport mechanisms, might be more efficient at directing molecules to specific targets or areas, enhancing communication and functional relationships.
High mutual information indicates a strong dependency on neurotransmitter concentration and ion channel states, meaning that changes in neurotransmitter concentration levels are effectively reflected in the state of the ion channels, which is crucial for accurate and rapid communication from the gut to the brain.
It demonstrates that the quantum communication model accurately captures the relationship between neurotransmitter concentration and ion channel states, validating the model's representation of synaptic communication.
Moreover, we have considered concentration to be the only input, and the channel is controlled by the phase generated by the $R_y$ gate. 
For rotation angle determination, achieving the maximum concentration on the VNM is essential. 
Hence, the VNM may remain aware of the incoming concentration flow of the neurotransmitters and adjust the angle rotation, which can support the concept of consciousness influenced by synaptic activity \bluecite{cook2008neuron}.

\section{Conclusion}
\label{sec:five}
In this study, we focused on the point-to-point channel on the VN membrane of GBAx and proposed a novel quantum circuit that resembles ion channel initiation by considering both passive diffusion and diffusion with active transport and only one vesicle in a single synapse. 
With the designed quantum circuit, we have simulated the VN membrane and obtained the output states. 
The desired output state, $|111\rangle$, indicates rapid opening of ion channels during the initial time frames, which facilitates fast action potential generation in the VN neuron. 
Furthermore, the QC channel through entropy and mutual information metrics are analyzed, which yields a maximum entropy of 0.8710 bits and a maximum MI of 0.6182 bits for passive diffusion, whereas a maximum entropy of 0.8885 bits and a maximum MI of 0.6609 bits for active transport. 
Higher entropy and MI in active diffusion compared to passive diffusion suggest that active transport processes introduce greater variability and more efficient information transfer within the system, indicating complex and highly regulated biological dynamics.
Overall, the main contribution of this paper is that it proposes a QC channel and verifies that such kind of quantum channel may exist at the synapse to rapidly transfer information and create an action potential on VN post-synaptic neuron. 
Moreover, our findings may suggest synapses can influence consciousness, as they show key awareness characteristics and adaptive responses to neurotransmitter concentration levels. 
This QC-based approach offers a novel framework for understanding synaptic processes and their computational replicas under the shade of quantum biology.

\bibliography{references}
\bibliographystyle{IEEEtran}
\end{document}